\documentclass[%
 aip,
 amsmath,amssymb,
 reprint,%
]{revtex4-1}

\usepackage{graphicx}
\usepackage{dcolumn}
\usepackage{bm}

\usepackage[utf8]{inputenc}
\usepackage[T1]{fontenc}
\usepackage{mathptmx}
\usepackage{etoolbox}
\usepackage{siunitx}

\makeatletter
\def\@email#1#2{%
 \endgroup
 \patchcmd{\titleblock@produce}
  {\frontmatter@RRAPformat}
  {\frontmatter@RRAPformat{\produce@RRAP{*#1\href{mailto:#2}{#2}}}\frontmatter@RRAPformat}
  {}{}
}%
\makeatother
\begin{document}

\preprint{AIP/123-QED}

\title{In situ calibration of microwave attenuation and gain using a cryogenic on-chip attenuator} 

\author{Thomas Descamps}
\affiliation{Department of Microtechnology and Nanoscience, Chalmers University of Technology, Gothenburg, Sweden.}
\email{thomas.descamps@chalmers.se}
\author{Linus Andersson}
\affiliation{Department of Microtechnology and Nanoscience, Chalmers University of Technology, Gothenburg, Sweden.}
\author{Vittorio Buccheri}
\affiliation{Department of Microtechnology and Nanoscience, Chalmers University of Technology, Gothenburg, Sweden.}
\author{Simon Sundelin}
\affiliation{Department of Microtechnology and Nanoscience, Chalmers University of Technology, Gothenburg, Sweden.}
\author{Mohammed Ali Aamir}
\author{Simone Gasparinetti}
\email{simoneg@chalmers.se}
\affiliation{Department of Microtechnology and Nanoscience, Chalmers University of Technology, Gothenburg, Sweden.}

\date{\today}

\begin{abstract}
Accurate in situ calibration of microwave attenuation and amplification-chain noise is essential for superconducting quantum circuits. We demonstrate a compact, self-calibrating cryogenic noise source based on an on-chip chromium attenuator that can be resistively heated with nanowatt-level power and directly integrated into a coaxial microwave line at the mixing-chamber stage. By comparing Johnson-Nyquist noise generated by Joule and microwave heating, measured through the amplification chain, the attenuation of the input line, and hence the gain of the chain, is determined without requiring knowledge of the attenuator temperature. The device exhibits millisecond-scale response times and negligible heating of the cryostat base plate. Using this approach, we determine the gain and added noise of a cryogenic amplification chain over the 4-8 GHz band. Our results provide a simple and accurate method to characterize near-quantum-limited parametric amplifiers used in superconducting-qubit readout.
\end{abstract}

\maketitle

Microwave measurements performed at cryogenic temperatures are essential across diverse low-temperature experiments, ranging from quantum information processing \cite{blais_circuit_2021, krantz_quantum_2019, krasnok_superconducting_2024, li_flying-qubit_2024, koolstra_coupling_2019, geng_high-fidelity_2025}, quantum optics \cite{you_atomic_2011, gu_microwave_2017}, quantum sensing \cite{danilin_quantum_2024, casariego_propagating_2023, teufel_nanomechanical_2009} to mesoscopic transport \cite{burkard_semiconductor_2023}. 
In these experiments, the device under test is typically driven through heavily attenuated microwave lines to suppress thermal noise \cite{krinner_engineering_2019}. It emits weak microwave signals, often at the single-photon level, that must be amplified during readout with low-noise high-electron-mobility transistor (HEMT) amplifiers \cite{montazeri_ultra-low-power_2016, zeng_sub-mw_2024, schleeh_ultralow-power_2012} and near-quantum-limited amplifiers, namely Josephson parametric amplifiers (JPAs) \cite{yamamoto_flux-driven_2008, castellanos-beltran_amplification_2008} and traveling-wave parametric amplifiers (TWPAs) \cite{macklin_nearquantum-limited_2015, vissers_low-noise_2016, malnou_three-wave_2021, ranadive_kerr_2022, ranadive_kerr_2022}.
Precise knowledge of the input-line attenuation, as well as the gain and noise temperature of the readout amplification chain, is therefore essential for quantitative interpretation of experiments and accurate characterization of amplifier performances.
Given the temperature dependence of these quantities, calibration must be performed in situ at cryogenic temperatures, preferably without modifying the experimental wiring in order to avoid errors from impedance mismatches and insertion losses.

Various calibration techniques have been developed to address these requirements. One approach exploits superconducting qubits either coupled to a cavity \cite{schuster_ac_2005} or to a transmission line to detect photons \cite{honigl-decrinis_two-level_2020} at the qubit transition frequency. Drive line attenuation as well as performance of the amplification chain referenced at the qubit chip can be measured with high precision, but only over a narrow frequency band, despite extension schemes up to 1 GHz with multilevel quantum systems \cite{kristen_amplitude_2020} or with flux-tunable qubits \cite{yan_distinguishing_2018}.
An alternative strategy employs bolometers consisting of a resistive absorber coupled to superconductor–normal metal–superconductor junctions embedded in an RF tank circuit, enabling broadband calibration of the attenuation of a microwave drive line \cite{girard_cryogenic_2023, singh_multiplexed_2025, gunyho_single-shot_2024}. However, in their current implementations, such devices cannot be readily inserted between the microwave drive line and the device under test without a significant increase in experimental complexity.
A third technique adapts the conventional room-temperature hot–cold noise method (Y-factor) to cryogenic environments \cite{heinz_cryogenic_2022, ardizzi_variable-temperature_2025, cano_ultra-wideband_2010}. Noise generated at two effective temperatures by a room temperature noise source is transmitted through the cryostat to a cold attenuator in front of the amplifier, and the resulting output noise is measured with a spectral analyzer to extract the amplifier noise temperature.
As a fourth approach, two main types of self-calibrated cryogenic noise sources have been reported: shot-noise tunnel junctions (SNTJs) \cite{malnou_low-noise_2024} and variable-temperature attenuators \cite{simbierowicz_characterizing_2021, simbierowicz_inherent_2024, goetz_photon_2017} or resistive loads \cite{malnou_low-noise_2024} equipped with a temperature sensor and a heater. Compared to the previous technique, these cryogenic noise sources can be mounted closer to the amplifier input plane. Variable-temperature attenuators can be assembled from commercial components and provide direct temperature readout; however, the heating power required to raise their temperature perturbs the cryostat base temperature, and their thermal response times are typically on the order of minutes. In contrast, SNTJs offer a larger dynamic range of output noise and nanosecond-scale response times, but they are sensitive to electrostatic discharge and rely on indirect temperature extraction through fitting procedures.

\begin{figure*}[ht]
\includegraphics[width=\textwidth]{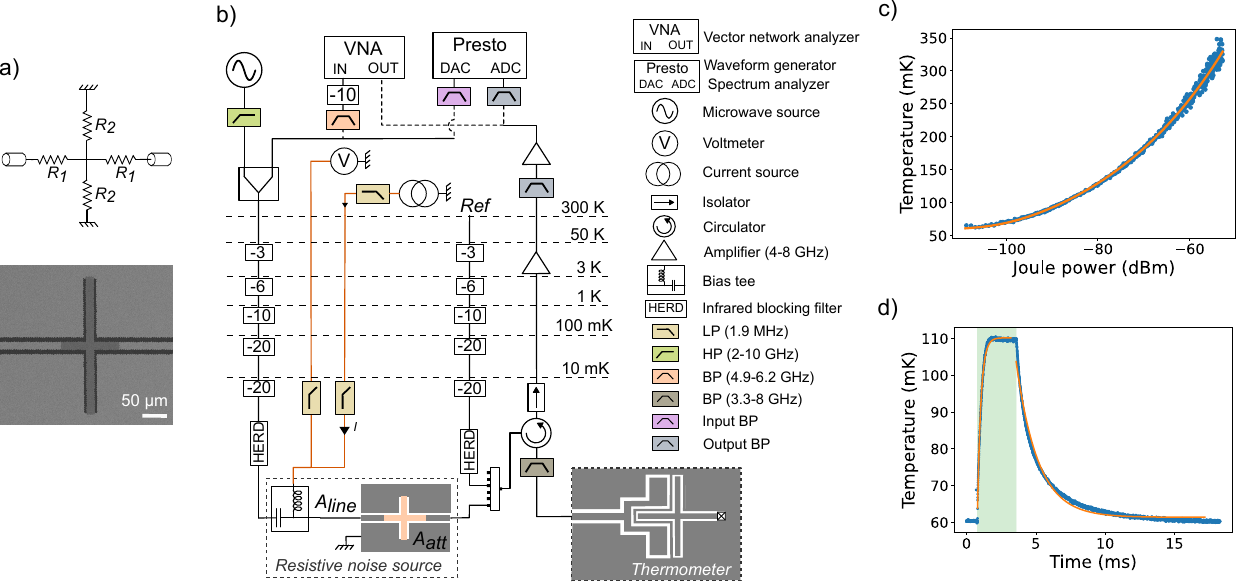}
\caption{\label{fig:Fig1}a) On-chip attenuator design with R$_{1} = 26$ $\Omega$ and R$_{2} = 70$ $\Omega$. b) Experimental setup showing the placement of the on-chip attenuator and the fixed-frequency radiation-field thermometer at the base plate of the dilution refrigerator, and their connection to microwave lines (black) and DC lines (orange). The input and output band-pass filters were chosen depending on the experiment. More details can be found in the main text. (BP: band-pass, LP: low-pass, HP: high-pass, $A_{\rm line}$: attenuation of the input line at the reference plane of the on-chip attenuator, $A_{\rm att}$: attenuation of the on-chip attenuator) c) The temperature of the 5.5 GHz microwave field measured with the thermometer as a function of the Joule power applied to the on-chip attenuator. The experimental data (blue) is fitted by the power law as shown in eq.~\ref{eq:PowerLaw} (orange). d) Thermalization times of the on-chip attenuator after applying a microwave heating pulse (green). The heating and cooling responses are fitted (orange) by single exponential to estimate the time constants.}
\end{figure*}

In this work, we demonstrate a compact cryogenic noise source based on an on-chip attenuator that can be heated using a dc current and directly integrated into a coaxial line at the mixing-chamber stage, upstream of the device under test. The attenuation of the drive line is obtained by comparing Johnson–Nyquist noise generated by Joule heating with that generated by microwave power dissipated in the attenuator \cite{simbierowicz_inherent_2024, girard_cryogenic_2023}. While previous works quantified noise using RF reflectometry with an LC tank circuit incorporating a bolometer, our approach directly measures the power spectral density of the generated noise using a room-temperature spectrum analyzer. The response time is on the order of 1 ms, and the locally generated heat does not measurably warm the cryostat base plate. With the attenuation thus determined, we quantify the gain and added noise of the HEMT-based amplification chain referenced to the output of the on-chip attenuator.

The self-calibrating cryogenic noise source using an on-chip microwave attenuator is shown in Fig.~\ref{fig:Fig1}(a). The attenuator is designed in a T-network configuration from a 60~nm chromium film (sheet resistance $R_{\rm s} = 8.7 \Omega/\square$ measured at 1.6~K) evaporated on a C-cut sapphire substrate and galvanically connected to 50-$\Omega$ niobium coplanar waveguides. Superconducting niobium is chosen for the leads for its perfect electrical conductivity while thermally insulating the chromium island from the rest of the circuitry \cite{feshchenko_thermal_2017}. The resistances of the attenuator \cite{yeh_microwave_2017} were selected to provide a nominal wide-band attenuation of $A_{\rm att}$ = -10~dB. The measured attenuation deviates by at most 1.5~dB in the 4–8 GHz band at 10~mK and varies by less than 0.15~dB between 10~mK and 3.8~K (Supplementary Information S1)

\begin{figure*}[ht]
\includegraphics[width=\textwidth]{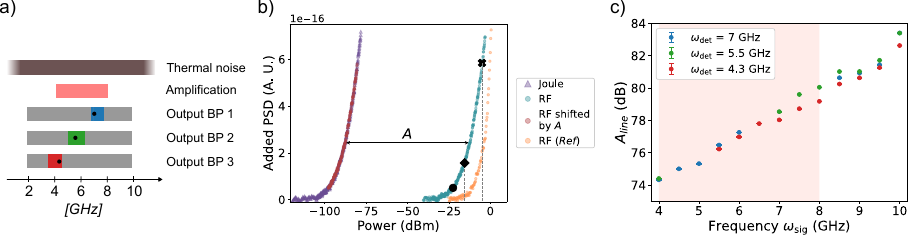}
\caption{\label{fig:Fig2} a) Sketch showing the frequency overlap between the thermal noise, the transmission band of the amplification chain, and the three output band-pass filters (1: 6.8-7.8~GHz, 2: 4.9-6.2~GHz; 3: 3.5-4.5~GHz) used to restrict the analog bandwidth in different power spectral density (PSD) measurements. For each filter, the signal tone to be calibrated, $\omega_{\rm sig}$, is swept in the rejection band of the filter (gray region), whereas noise is detected at $\omega_{\rm det}$ (black dot) in the transmission band (colored region). b) Power spectral densities measured at $\omega_{\rm det}=7$~GHz by Joule heating (purple) and a RF heating (blue) at $\omega_{\rm sig}=4$~GHz of the on-chip attenuator. The tone applied was rejected at room temperature with the output band-pass filter 1. The Joule and RF heating traces are separated by the attenuation $A$ of the input line. The tone heating was also acquired via the reference line without the on-chip attenuator (orange) to show the onset of heating of the bulkhead attenuators (diamond marker). Since their contribution is negligible (see main text), the best overlap (red) of the RF heating with the Joule heating giving the attenuation $A$ was fitted from $P_{\rm sig}=-22.5$~dBm (above the noise floor, dot marker) until an arbitrary power $P_{\rm sig}=-5$~dBm (cross marker). c) Attenuation profile obtained by comparing Joule and RF heating at different $\omega_{\rm sig}$. Three detection frequencies were considered and implemented with the set of output band-passes depicted in a). The red band in the background highlights the bandwidth of the amplification chain.}
\end{figure*}

The input of the on-chip attenuator, mounted at the base plate of the dilution refrigerator [Fig.~\ref{fig:Fig1}(b)], is connected to a commercially available bias tee, enabling simultaneous transmission of RF signals and DC connectivity in a four-probe configuration. The input RF coaxial line is attenuated at multiple temperature stages by bulkhead attenuators and filtered at the base plate using a high-energy radiation drain (HERD) filter~\cite{andersson_co-designed_2025} to suppress blackbody radiation and electrical noise, thereby minimizing undesired heating of the on-chip attenuator. The output of the on-chip attenuator is connected to a microwave switch, along with a reference input line having the same nominal attenuation but without the bias tee or the on-chip attenuator. The output port of the switch is routed via a circulator and a band-pass filter (3.3-8~GHz) to a thermometer consisting of a fixed-frequency superconducting transmon (5.5~GHz) strongly coupled to the end of a transmission line~\cite{scigliuzzo_primary_2020} (other parameters in Supplementary Information S2). This thermometer is used solely to cross-check the performance of the noise source and is not required for its operation. The reflected signal is then routed through one isolator and amplified by a low-noise high-electron-mobility transistor (HEMT) amplifier at the 3~K stage, followed by two cascaded room-temperature amplifiers. The DC wiring comprises thermocoaxial cables filtered at the base plate by reactive low-pass filters (1.9~MHz). By applying a DC current through the on-chip attenuator to ground,  we measure the resistance of its active region using a four-probe technique, obtaining $R_{\rm att} = 67$ $\Omega$.
Based on the series of bulkhead attenuators along the input line, typical for superconducting qubit experiments, we predict a minimum achievable effective base temperature of 44~mK at 5.5~GHz using the beam-splitter model of an attenuator for thermal photons \cite{krinner_engineering_2019}. By measuring the real part of the reflectance at the thermometer resonance with a vector network analyzer (Supplementary Information S3), we infer a radiation field temperature of $48.6 \pm 0.3$~mK for the reference line, validating good thermalization of the coaxial line and bulkhead attenuators. When the on-chip attenuator is inserted without external driving, the inferred temperature increases to $60.4 \pm 0.2$~mK, indicating unintentional heating of the on-chip attenuator due to residual unfiltered electrical noise.

The electron temperature $T_{\rm e}$ in the chromium film can be increased by applying Joule power $P_{\rm Joule}$ to the on-chip attenuator. Assuming no power dissipation in the superconducting Nb film, this external power is converted to electron-phonon scattering in the chromium film, described by the power balance at steady state

\begin{equation}\label{eq:PowerLaw}
    P_{\rm Joule} = R_{\rm att}I^{2} = \Sigma_{\rm Cr}V(T_{e}^{\alpha} - T_{0}^{\alpha}),
\end{equation}

where $V$ is the volume of the chromium attenuator, $\Sigma_{\rm Cr}$ the electron-phonon coupling and $\alpha$ a power exponent. By fitting the temperature obtained in Fig.~\ref{fig:Fig1}(c), we find $\Sigma_{\rm Cr} = 2.53\times10^{10} \pm 6.7\times10^{8}$ Wm$^{-3}$K$^{-\alpha}$ and $\alpha = 6.72 \pm 0.015$, comparable to values reported for chromium films on Silicon \cite{subero_bolometric_2023}. We note that the maximum temperature reached is bounded by the operating range of the (qubit-based) thermometer and not by the heating range of the on-chip attenuator. Since only 1~nW of dissipated power is required to generate a thermal field equivalent to 257~mK, this device can be safely operated without causing undesirable heating of the base plate. 

The thermal response times of the on-chip attenuator are characterized by applying a microwave pulse at 9~GHz to heat it, in addition to the thermometer drive tone at 5.5~GHz. The 9-GHz pulse is rejected by the band-pass filter preceding the thermometer, while the signal reflected at 5.5~GHz is recorded with the digitizer. The amplitude of the pulse was set to raise the temperature to 110~mK, and the resulting heating and cooling transients are shown in Fig.~\ref{fig:Fig1}(d). The response times are governed by the ratio of the heat capacity to the thermal conductance of the chromium film, both of which are temperature-dependent \cite{gasparinetti_fast_2015}. Since the heat capacity is not independently known, the transient responses are approximated by a single exponential and fitted to the full dataset, yielding time constants of 0.25 and 1.57~ms for heating and cooling, respectively. These settling times are orders of magnitude faster than those reported for variable-temperature noise sources based on commercial bulkhead attenuators, which are typically on the order of minutes~\cite{simbierowicz_characterizing_2021}. With this improvement, the noise characterization of low-noise cryogenic amplifiers can be substantially sped up.

The attenuation at the input plane of the on-chip attenuator, $A_{\rm line}$, is quantified by comparing the output noise power density (PSD) generated either by Joule heating of the on-chip attenuator or by heating induced by a microwave tone at frequency $\omega_{\rm sig}$ transmitted through it (referred to as RF heating). In particular, the PSD is measured using a spectrum analyzer at the output of the amplification chain, and the value at a single frequency $\omega_{\rm det}$ is obtained by integration over a 5~MHz span with a resolution bandwidth of 3052~Hz. The microwave heating tone is rejected by two identical room-temperature band-pass filters placed before the spectrum analyzer and the thermal noise generated is measured at $\omega_{\rm det}$ within the filter pass-band [Fig. ~\ref{fig:Fig2}(a)]. Although the Johnson–Nyquist noise depends on the frequency at which it is generated~\cite{clerk_introduction_2010}, we assume that it can be detected at an arbitrary frequency $\omega_{\rm det}$ to compare noise generated by Joule and RF heating. Every PSD acquisition is interleaved between heating on and off in order to extract the added PSD. 
A calibrated signal generator is used to generate the heating tone with output power $P_{\rm sig}$, resulting in a microwave power dissipated in the on-chip attenuator given by $P_{\rm RF} = A_{\rm line}(1-A_{\rm att})P_{\rm sig}$. As shown in Fig.~\ref{fig:Fig2}(b), the PSD traces obtained by increasing either $P_{\rm sig}$ or $P_{\rm Joule}$ follow the same trend and differ only by a horizontal shift in power corresponding to the attenuation factor $A = A_{\rm line}(1-A_{\rm att})$.

To assess the contribution of the bulkhead attenuators to the noise generated, we drive the RF tone through the reference line nominally identical but which does not include the on-chip attenuator. In this case, the onset of the noise generated by RF heating occurs at higher power, indicating that measurable noise is generated predominantly by the on-chip attenuator, which is less efficiently thermalized than the preceding bulkhead attenuators. We extract $A_{\mathrm{line}}$ from the horizontal shift required to align the PSD trace obtained from RF heating with that obtained from Joule heating [Fig.~\ref{fig:Fig2}(b)]. We optimize the shift using a least-squares method over the microwave power range where the generated noise lies sufficiently above the noise floor ($P_{\rm sig} > -22.5$~dBm, dot marker) and originates solely from the on-chip attenuator ($P_{\rm sig} < -15.5$~dBm, diamond marker). For a microwave tone at $\omega_{\rm sig}=4$~GHz, we obtain $A_{\rm line}=-74.2\pm0.02$~dB by measuring the PSD at $\omega_{\rm det}=7$~GHz. 
While accurate, this approach relies on identifying the onset of heating in the bulkhead attenuators. In practice, this requires a nominally identical reference line, making the method experimentally inconvenient. At higher signal powers, the total measured PSD, $\text{PSD}_{\rm tot}$, contains contributions from both the on-chip attenuator, $\text{PSD}_{\rm NS}$, and the bulkhead attenuators, $\text{PSD}_{\rm other}$, attenuated by $A_{\rm att}$: $\text{PSD}_{\rm tot} = \text{PSD}_{\rm NS} + A_{\rm att}\times \text{PSD}_{\rm other}$. Using the reference line, we find that $\text{PSD}_{\rm other} = 0.34\times \text{PSD}_{\rm tot}$ at $P_{\rm sig}=-5$~dBm, indicating that the on-chip attenuator contributes to $\approx$ 97\% of the total noise generated. By overlapping the PSD obtained from Joule heating and RF heating for tone powers up $P_{\rm sig}=-5$~dBm (cross marker), we extract $A_{\rm line}=74.3\pm0.01$~dBm, which indicates a discrepancy of 0.2~dB with the previously obtained attenuation. This confirms that, in this power range, the contribution of noise generated by the bulkhead attenuators can be neglected, and all attenuation estimation was carried out until $P_{\rm sig}=-5$~dBm.
Using this method, the attenuation profile of the input line can be obtained by sweeping $\omega_{\rm sig}$ (Fig.~\ref{fig:Fig2}(c)). The band-pass filter used to reject the heating tone before the spectrum analyzer, and thus the acquisition frequency $\omega_{\rm det}$, was selected accordingly. At frequencies $\omega_{\rm sig}$ where the band-pass regions overlap, the attenuations measured agree within at most 1~dB, supporting the assumption that the attenuation calibration can be made at an arbitrary detection frequency $\omega_{\rm det}$. In the following, we therefore assign a 0.5~dB uncertainty to the attenuation.
To validate the method, we compare the attenuation of $76.3 \pm 1$~dB measured at 5.5~GHz with an independent calibration using the temperature sensor (Supplementary Information S4). The latter yields an attenuation of $89.3 \pm 0.025$~dB referenced to the sensor input, corresponding to an additional loss of 13.1 dB between the input of the on-chip attenuator and the sensor. This excess loss indicates 2.3~dB of cabling attenuation combined with the independently measured 10.8~dB attenuation of the on-chip attenuator.

\begin{figure}[ht]
\includegraphics[width=\linewidth]{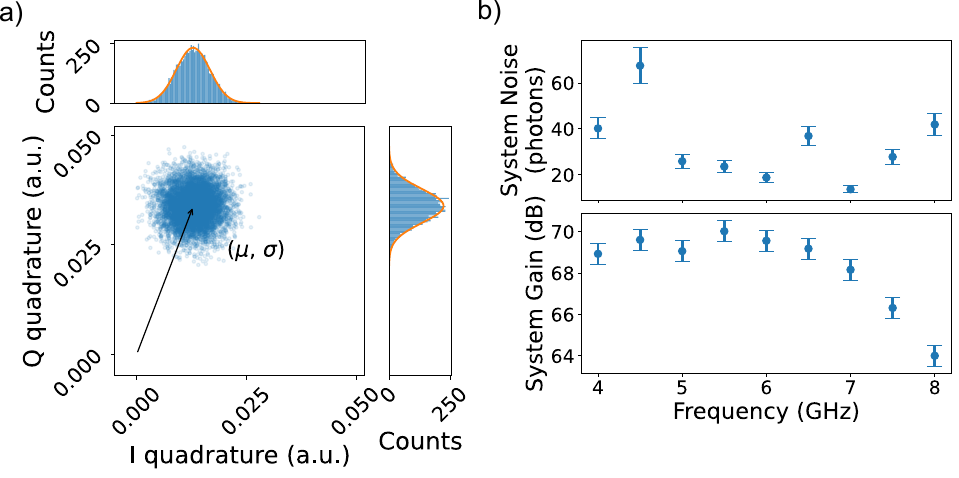}
\caption{\label{fig:Fig3}a) Scattering in the $I-Q$ plane of a weak coherent tone transmitted through the amplification chain. The projections onto the $I$ and $Q$ axes are fitted with Gaussian distributions, from which the mean $\mu$ and standard deviation $\sigma$ are extracted. b) Added noise (top) and gain (bottom) of the amplification chain, referenced  to the input plane of the noise source as mounted in Fig.~\ref{fig:Fig1}(b). The error bars are obtained by propagating the uncertainty in the attenuation, taken to be $\pm 0.5$~dB.}
\end{figure}

With the attenuation up to the noise source known, the system noise of the amplification chain which includes all components after the on-chip attenuator can be determined by measuring the scattering of a transmitted weak coherent tone in the $I-Q$ quadrature plane over an integration time $t_{\rm int}$ [Fig.\ref{fig:Fig3}(a)], as typically done in qubit readout \cite{blais_circuit_2021}. The mean displacement $\mu$ and the standard deviation $\sigma$ of the resulting distribution are related to the added noise of the amplification chain, $n_{\rm add}$, by

\begin{equation}\label{eq:added_noise}
    n_{\rm add} = \frac{N_{\rm sig}}{\text{SNR}^{2}} - \frac{1}{2} = \frac{P_{\rm input}t_{\rm int}}{\hbar\omega_{\rm sig}}\frac{\sigma^{2}}{\mu^{2}} - \frac{1}{2},
\end{equation}

where $N_{\rm sig}$ is the number of photons at frequency $\omega_{\rm sig}$ at the input of the amplification chain, $\text{SNR}$ the signal to noise ratio and $P_{\rm input}=A_{\rm line}A_{\rm att}P_{\rm sig}$ the power at the input of the amplification chain.
The tone power $P_{\rm sig} = -28$ dBm is chosen sufficiently low to avoid heating the on-chip attenuator. By sweeping the tone frequency, we extract the system noise [Fig.~\ref{fig:Fig3}(b)], obtaining $23.6 \pm 2.8$ photons at 5.5~GHz for an overall gain of $70.2 \pm 0.5$~dB. We emphasize that the system noise found does not represent the intrinsic added noise of the HEMT ($\approx$ 5 photons at 5~K according to the datasheet), since the reference plane is taken at the output of the on-chip attenuator rather than at the HEMT input. This limitation can be removed by directly connecting the output of the on-chip attenuator to the isolator. For comparison, an independent calibration using the temperature sensor referenced to its input plane yields $14.2 \pm 2.3$ photons (Supplementary Information S4).

In conclusion, we demonstrated a compact cryogenic noise source based on a variable-temperature chromium attenuator requiring nanowatt-level heating power and exhibiting millisecond-scale response times. By comparing power spectrum density generated by Joule and microwave heating, the attenuation of a microwave drive line is determined with an accuracy of at most 0.5~dB without requiring knowledge of the attenuator temperature. The added noise of the amplification chain, referenced from the output of the on-chip attenuation, was then found from the scattering of a weak coherent tone. The device can be readily integrated into experiments for calibrating drive-line attenuation or characterizing the gain and added noise of low-noise cryogenic amplifiers or quantum-limited parametric amplifiers. Future versions could further reduce the footprint by integrating the required bias tee on chip.

\begin{acknowledgments}

This work was funded by the European Union’s HORIZON-RIA Programme under Grant No.~101135240 JOGATE.
M.A.A. and S.S.~acknowledge financial support from the European Union via Grant No.~101080167 ASPECTS.
S.G.~acknowledges financial support from the European Research Council (Grant No.~101041744 ESQuAT) and the Knut and Alice Wallenberg Foundation through the Wallenberg Centre for Quantum Technology (WACQT).
The devices used in this work were fabricated in the Chalmers Myfab cleanroom facility.

External interest disclosure: S.G.~and L.A.~are co-founders and equity holders in Sweden Quantum AB.

\end{acknowledgments}

\bibliography{BIB_NoiseSource}

\clearpage
\onecolumngrid

\begin{center}
\textbf{\large Supplementary Information}
\end{center}

\setcounter{figure}{0}
\renewcommand{\thefigure}{S\arabic{figure}}
\setcounter{section}{0}
\renewcommand{\thesection}{S\arabic{section}}

\section{Cryogenic characterization of the on-chip attenuation}

\begin{figure}[ht]
\includegraphics[width=0.8\linewidth]{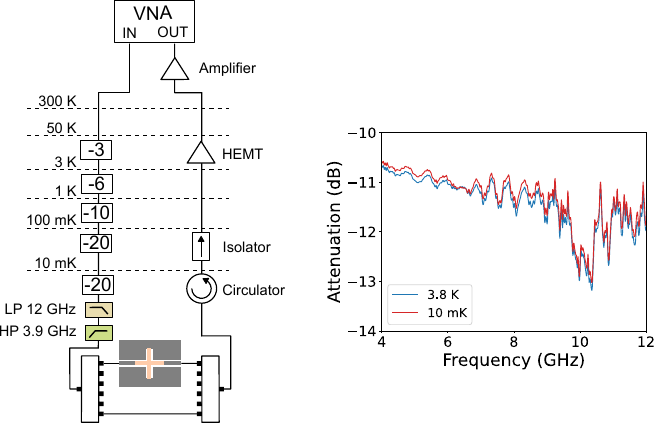}
\caption{\label{fig:S1}}
\end{figure}

To characterize the scattering parameters of the on-chip attenuator, the device is mounted between two mechanical switches (Radiall) at the mixing chamber of the dilution refrigerator. The attenuation was found by comparing the transmission coefficient $S_{21}$ to a through. Over the 4–12 GHz frequency range, the measured attenuation is bounded between -10.6 and -13 dB (Fig.~\ref{fig:S1}).

\section{Parameters of the radiation field thermometer}

\begin{table}[ht]
\begin{ruledtabular}
\begin{tabular}{cc}
Parameter &Value \\
\hline
Fundamental resonance frequency & $\omega_{\rm ge}/2\pi=5.496$ GHz \\
Anharmonicity  & $\alpha/2\pi=-200$ MHz \\
Linewidth  & $\Gamma/2\pi=35.8$ MHz\\
\end{tabular}
\end{ruledtabular}
\caption{\label{fig:table1}}
\end{table}

The frequency, the anharmonicity and the linewidth of the radiation-field thermometer (Table~\ref{fig:table1}) were obtained with the VNA by measuring the reflection $|\rm Re[r(\omega)]|$, as discussed in ref.\cite{scigliuzzo_primary_2020}.

\section{Temperature fitting}

\begin{figure}[ht]
\includegraphics[width=0.5\linewidth]{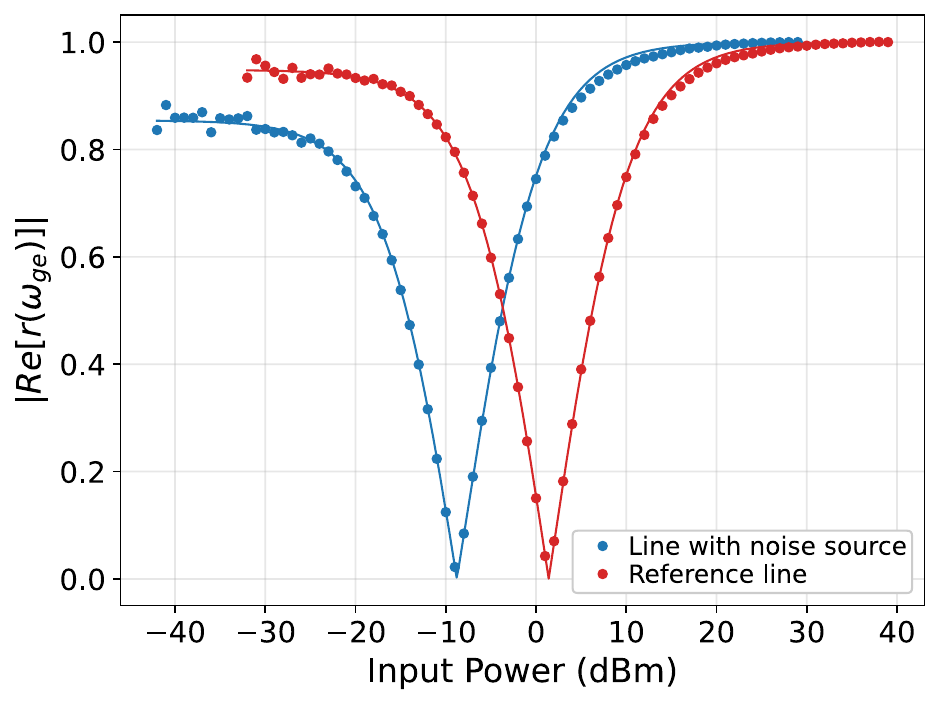}
\caption{\label{fig:S2}}
\end{figure}

The system temperature is characterized by measuring the reflection $|\rm Re[r(\omega_{\rm ge})]|$ of a microwave probe tone at the thermometer resonance $\omega_{\rm ge}$ using a vector network analyzer (VNA) as a function of the output power $P_{\rm VNA}$. To access the relevant power regime, external attenuators were placed directly at the output of the VNA, resulting in an input power $P_{\rm in} = P_{\rm VNA} - 16$ for the line containing the noise source and $P_{\rm in} = P_{\rm VNA} - 26$ for the reference line.
As illustrated in Fig.~\ref{fig:S2}, the power-dependent responses exhibit a characteristic V-shaped profile, attributed to destructive interference between the emitter's scattered field and the phase-inverted reflection at the waveguide boundary. In the low-power limit, the reflection magnitude $|Re[r(\omega_{\rm ge})]|$ converges to non-unity values due to thermal occupation $n_{\rm r}$ in the waveguide. By fitting with the model described in ref.\cite{scigliuzzo_primary_2020} (Appendix D) for a 3-level system, we extract $n_{\rm r}^{\rm NS} = 0.013 \pm 1.8\times10^{-4}$ and $n_{\rm r}^{\rm ref} = 0.0044 \pm 1.3\times10^{-4}$, which correspond to temperatures of $T^{\rm NS} = 60.4 \pm 0.2$ mK and $T^{\rm ref} = 48.6 \pm 0.3$ mK according to Bose-Einstein distribution.

\section{Attenuation and gain obtained with the radiation field thermometer}

The attenuation $A$ of the input line can be determined at the temperature sensor resonance frequency $\omega_{\rm ge}$, as detailed in ref. \cite{scigliuzzo_primary_2020}:

\begin{equation}\label{eq:PowerLaw}
    A = \frac{\hbar \omega_{\rm ge}\Omega^{2}}{4\Gamma P_{\rm in}},
\end{equation}

where $P_{\rm in}$ the input power and $\Omega$ the drive rate. The reflected microwave signal  $|Re[r(\omega_{\rm ge})]|$ reaches a minimum for a certain power $P_{\rm min} = -8.4 \pm 0.032$ dBm, at which the drive rate relates to the linewidth as $\Omega/\Gamma \approx 1/\sqrt2$. Thus, the attenuation obtained is $A = 89.3 \pm 0.025$ dB at the reference plane of the temperature sensor. The discrepancy of 13.0 dB compared to the on-chip attenuator reference plane is consistent with the attenuation of the absorber (measured 10.8 dB) and supplementary cable losses estimated at 2.2 dB. 
By repeating the same noise measurement at 5.5 GHz as in the main text, we obtain the noise figure of $14.2 \pm 2.3$ photons. 

\end{document}